\documentclass[runningheads]{llncs}

\usepackage{graphicx}
\usepackage{booktabs}
\usepackage{tabularx}
\usepackage{hyperref}
\usepackage{paralist}
\usepackage{tabulary}
\usepackage{ragged2e}
\usepackage{subcaption}
\usepackage{wrapfig}
\usepackage{cite}
\usepackage{comment}
\usepackage{amsmath}
\usepackage{cancel}
\usepackage{xcolor}
\usepackage{paralist}
\usepackage{caption}
\usepackage{tikz}

\newcolumntype{Y}{>{\centering\arraybackslash}X}
\newcommand{\mypar}[1]{\vspace{0.1em}\noindent\textbf{#1.}}
\newcommand{\myparit}[1]{\vspace{0.1em}\noindent\textit{#1.}}

\begin{document}

\title{Identifying Process Improvement Opportunities through Process Execution Benchmarking}
\titlerunning{Process Execution Benchmarking}

\author{Luka Abb\orcidID{0000-0002-4263-8438}\inst{1} 
\and Majid Rafiei\orcidID{0000-0001-7161-6927}\inst{2}
\and Timotheus Kampik\orcidID{0000-0002-6458-2252}\inst{2}
\and Jana-Rebecca Rehse\orcidID{0000-0001-5707-6944}\inst{1}}

\institute{%
  University of Mannheim, Germany \\ 
  \email{\{luka.abb, rehse\}@uni-mannheim.de} \\
  \and
  SAP Signavio, Berlin, Germany \\ 
  \email{\{majid.rafiei, timotheus.kampik\}@sap.com}
}

\maketitle              
\begin{abstract}
Benchmarking functionalities in current commercial process mining tools allow organizations to contextualize their process performance through high-level performance indicators, such as completion rate or throughput time. However, they do not suggest any measures to close potential performance gaps. To address this limitation, we propose a prescriptive technique for process execution benchmarking that recommends targeted process changes to improve process performance. The technique compares an event log from an ``own'' process to one from a selected benchmark process to identify potential activity replacements, based on behavioral similarity. It then evaluates each proposed change in terms of its feasibility and its estimated performance impact. The result is a list of potential process modifications that can serve as evidence-based decision support for process improvement initiatives.

\end{abstract}

\section{Introduction}

Benchmarking is a well-established method for organizations to assess their performance relative to peers, set performance goals, and identify areas for operational improvement \cite{benchmarking_general}. 
In particular, benchmarking often focuses on business processes \cite{process_benchmarking}.
For example, a finance organization might benchmark its purchase-to-pay (P2P) workflow completion rates by comparing them to those of other organizations that use the same standard software, thus facilitating comparability. 
This comparison allows the organization to contextualize its own completion rate, determining whether its performance is ``good'' (significantly above the median) or ``bad'' (below average, based on the overall distribution).

To support such analyses, commercial process mining tools offer benchmarking capabilities based on process-level performance indicators, such as completion rate, automation rate, or average throughput time \cite{celonisweb,sapweb}. These indicators tell organizations how well their processes run relative to those of their competitors. However, they do not offer insights on what process improvements would be necessary to match the performance of industry peers.

To address this limitation, it is necessary to move from a high-level comparison of performance indicators to a detailed comparison of the actual business process \emph{execution} \cite{process_benchmarking}. 
This approach, which we call \emph{process execution benchmarking}, involves the in-depth comparison of two or more event logs that record executions of processes sharing relevant properties, e.g., the process type. 
Process execution benchmarking can be conducted externally, by comparing process executions with those of similar organizations, or internally, by analyzing event logs from different units within the same company.
This latter approach holds significant potential for process improvement: In large, geographically dispersed organizations, operational knowledge often remains isolated within specific divisions or teams \cite{knowledge_sharing1,knowledge_sharing2}. 
This can lead to inefficiencies, as one organizational branch might excel in a particular process but lack the means or incentives to communicate and share their expertise with others \cite{knowledge_sharing3}. 
Execution benchmarking addresses this in a data-driven way. 
For instance, the finance organization could use it to compare the P2P processes in its different country-level branches. By studying the process variants frequently executed in the best-performing branch, it can determine best practices and implement them in other branches as well.

Previous work on comparative analysis of event logs primarily focused on descriptive insights, aiming to uncover and visualize differences in control-flow and performance \cite{process_variants_review}. In this paper, we instead take a \emph{prescriptive} approach: We introduce a technique that takes an ``own'' and a ``benchmark'' event log and identifies actionable \cite{Park2022_actionable,Stein2024_improvement} process changes expected to improve process performance.
It outputs a set of behaviorally plausible process modifications, each associated with measures for feasibility and estimated performance impact. The potential modifications can then be sorted, filtered, and compiled into a list of recommended process changes that is provided to a process manager.

Specifically, we concentrate on process changes in the form of activity replacements. This is motivated by the fact that activity selection is critical for process optimization \cite{Scheer2015}: In many information systems, e.g., ERP systems, the same process can be executed by multiple configurations or pathways. Hence, different organizations or branches often implement different approaches for key process steps, but may not be aware of the impact that their activity-level choices have on their overall process performance. Our technique is meant to make these performance implications visible and provide concrete improvement suggestions based on practices that are already successfully employed by the benchmark.

\section{Problem Illustration}
\label{sec:running_example}

\enlargethispage{\baselineskip}
To illustrate our technique, we use a standardized purchasing process in two versions, shown in \autoref{fig:example_models}: The ``own'' version to be improved and the ``benchmark'' version, from which we draw potential improvements. 
The process models are for illustration only; our technique does not rely on models. 
Both process versions include essentially the same four steps: Create a purchase order (PO), assign it to a purchase requisition (PR), release the PO, and post an invoice receipt. However, the first and third step are implemented in different ways. 
\autoref{tab:example_logs} shows minimal event logs with the execution variants these models permit.

\begin{minipage}{\linewidth}
    \centering
    \vspace{1.25em}
    
    \includegraphics[width=1\textwidth]{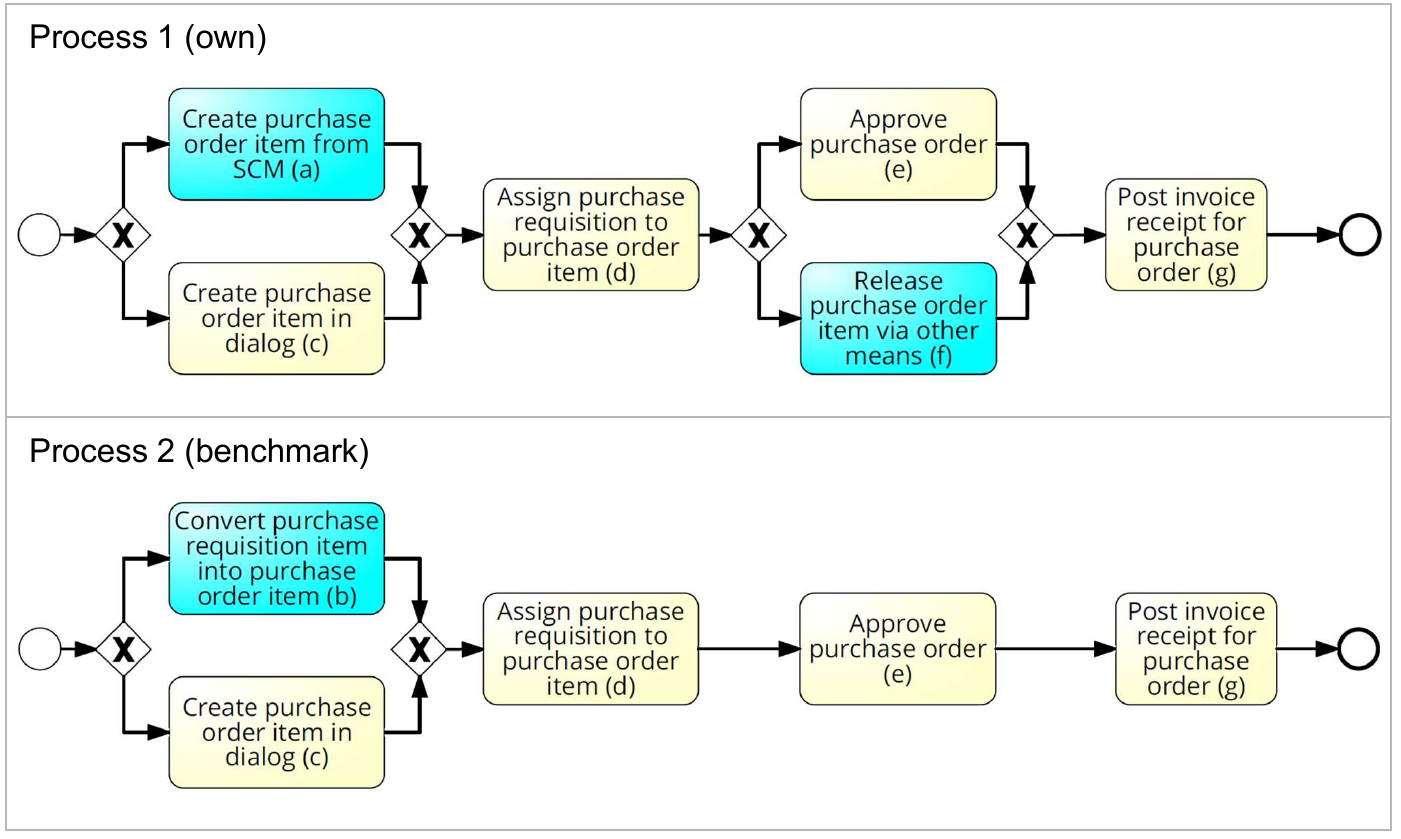}
    \captionof{figure}{Running example own and benchmark versions of a (simplified) purchasing process in an SAP ERP system. Disparities are highlighted in blue.}
    \label{fig:example_models}

    \vspace{0.25em}
    
    \centering
    \captionof{table}{Running example event logs with all execution variants that the respective process models permit.}
    \vspace{0.5em}
    \begin{subtable}[t]{0.45\textwidth}
        \centering
        \begin{tabular}{c}
            \toprule
            $L_1$ \\ \midrule
            $\langle a, d, e, g \rangle$,
            $\langle a, d, f, g \rangle$ \\
            $\langle c, d, e, g \rangle$,
            $\langle c, d, f, g \rangle$ \\
            \bottomrule
        \end{tabular}
        \caption{Event log $L_1$.}
        \label{tab:event_log_L1}
    \end{subtable}%
    \hfill
    \begin{subtable}[t]{0.45\textwidth}
        \centering
        \begin{tabular}{c}
            \toprule
            $L_2$ \\ \midrule
            $\langle b, d, e, g \rangle$,
            $\langle c, d, e, g \rangle$ \\
            \\
            \bottomrule
        \end{tabular}
        \caption{Event log $L_2$.}
        \label{tab:event_log_L2}
    \end{subtable}
    
    \label{tab:example_logs}
    \vspace{1em}
\end{minipage}

\noindent
This example showcases the two modifications that our approach should identify:

\begin{compactenum}[(1)]
    \item \textbf{True replacement:} In the own process, a PO can be created from the Supply Chain Management (SCM) system. The benchmark process instead allows PO items to be created by converting items from a PR. Activity $b$ is hence a possible replacement for activity $a$ that could lead to faster process execution, since the PO items do not have to be generated from scratch.
    \item \textbf{Choice streamlining:} In the own process, a PO can be released by approval or through other means (e.g., directly, without involving central procurement), consolidated in activity $f$. The benchmark process always requires approval, which could, e.g., lead to better compliance. 
    On a process level, this simplifies the PO release, but on a log level, it can be interpreted as a replacement, since every execution variant in the own log that would execute activity $f$ at this point will instead contain activity $e$ in the benchmark log.
\end{compactenum}

In the following, we will use this example to illustrate how our technique finds potential improvements to the process.

\section{Foundations}
\label{sec:behavioral_relations}

\mypar{Event Logs}
Process mining relies on event logs, i.e., collections of data records that capture the execution of processes in organizations. An event log consists of multiple unique cases, where each case represents one complete execution of a process. A case is composed of a sequence of unique events, called a trace. An event is a record of an activity that occurred during the execution of the process.

Formally, an \textit{event log} $L$ is a set of traces, denoted by
$L = \{ t_1, t_2, \dots, t_n \}$. Each $t_i, 1 \leq i \leq n$ is a trace, and  $n = |L|$ is the number of traces in the event log. A \textit{trace} $t$ is a sequence (totally ordered set) of events corresponding to a single process execution (a case). A trace is denoted as $t = \langle e_1, e_2, \dots, e_m \rangle$,
where each $e_i$ is an event, $m$ is the number of events in the trace, and for $e_j, e_k$, $1 \leq j < k \leq m$, we say that ``$e_j$ occurs before $e_k$ (in $t$)''.
An \textit{event} $e = (c, a)$ is a tuple of attributes, with at least two components, where $c$ is a unique identifier for the case to which the event belongs and $a$ is the name of the executed activity. The set of all activities in the event log is denoted as $A$.
A \textit{variant} $v \in [\langle a_1, a_2, \dots, a_m \rangle \mid \langle (c_1,a_1), (c_1,a_2), \dots, (c_1,a_m) \rangle \in L]$ is a unique sequence of activity attributes of events that appear in a trace. 
$V$ denotes a set of variants and $T(v)$ denotes the set of all traces that exhibit a variant $v \in V$.


\mypar{Ordering Patterns and Behavioral Relations}
Within a trace $t$, the relative ordering of two activities $a, b \in A$ can be described by the ordering pattern $a >_t b$ if $a$ is executed at some point before $b$ (i.e., $b$ eventually follows $a$) and $a \not>_t b$ otherwise. 
Based on these patterns, any pair of activities $a, b \in A$ in the event log can be characterized by one of three behavioral relations \cite{alpha_miner,Weidlich2011_behavioral_relations}:

\myparit{Strict Order ($a \rightarrow b$)} 
Activity $a$ occurs before $b$ in all traces where both occur:
\vspace{-0.05em}
    $$ \forall \: t \in L: \: a >_t b \: \text{and} \: b \not>_t a $$

\noindent A reverse strict order relationship is written as $a \leftarrow b$.
    
\myparit{Exclusiveness ($a \# b$)} 
Activities $a$ and $b$ never occur together in any trace:
\vspace{-0.05em}
    $$ \forall \: t \in L: \: a \not>_t b \: \text{and} \: b \not>_t a   $$
    
\myparit{Interleaving Order ($a \parallel b$)} 
Activities $a$ and $b$ can occur in any order within a trace, i.e., in at least one trace $a$ occurs before $b$ and vice versa:
\vspace{-0.05em}
    $$ \exists \: t \in L \: | \: a >_t b \quad \text{and} \quad \exists \: t \in L \: | \: b >_t a $$

\noindent Per these definitions, an activity has an interleaving relation to itself if it can repeat within a trace (i.e., is part of a loop) and an exclusive one if it cannot.

\section{Technique}

\begin{figure}[tb]
    \centering
    \includegraphics[width=1\linewidth]{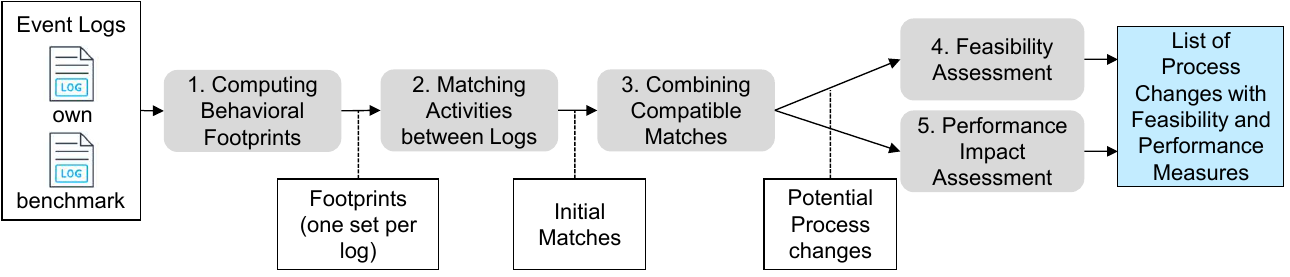}
    \caption{High-level overview of the proposed technique and intermediary results}
    \label{fig:overview}
    \vspace{-1.0em}
\end{figure}

Our technique aims to provide a process manager with a list of process changes that could improve process performance. It requires as input two event logs, $L_1$ and $L_2$, which capture executions of the own and the benchmark process, respectively. Its output is a list of behaviorally plausible process modifications, each associated with a feasibility and a performance assessment.
To identify these modifications, we establish behavioral footprints per activity in both logs (\autoref{sec:footprints}) and use these to identify initial activity ``matches'', i.e. behaviorally plausible replacement options (\autoref{sec:matching}). 
Replacement options that can be implemented together are grouped into sets, representing potential process changes (\autoref{sec:compatibility}). Finally, we assess each process change with regard to feasibility (likelihood to result in a valid process) (\autoref{sec:feasibility_assessment}) and expected performance impact (\autoref{sec:performance_assessment}). 
These five steps are summarized in \autoref{fig:overview}.

We designed this technique based on the following assumptions:

\begin{compactenum}[(1)]
    \item \textit{Same process type:} Both event logs capture the same process type, for example, a purchase-to-pay process. Otherwise, it would hardly be possible to improve one process using data-driven insights from the other.
    \item \textit{Standardized activity names:} Activity names are standardized, i.e., two activities in $L_1$ and $L_2$ with the same name refer to the same process step and two activities with different names refer to different process steps. This assumption is realistic in environments where organizations adopt enterprise information systems from a standard software and implement the standardized reference process models that this software contains. This holds, for example, in the SAP environment, where many different organizations implement standardized processes according to the SAP reference model \cite{curran1997sap}.
    \item \textit{Common performance measure:} Both event logs include at least one common case-level performance measure, for example, a successful completion indicator, the throughput time, or a customer satisfaction score.
\end{compactenum} 
\noindent
In the following, we describe the steps of our technique in more detail.

\subsection{Computing Behavioral Footprints}
\label{sec:footprints}

The first two steps in our technique aim to establish a set of potential activity replacements that make sense from a behavioral perspective. Intuitively, we only want to replace an activity with one that implements an analogous step in the process, i.e., accomplishes essentially the same advancement towards the process goal. In the running example, it is easy to see that replacing $a$ (creating a PO) with $e$ (releasing a PO) is not a valid change because the latter is always executed after the former in the own process. Replacing $a$ with $b$, however, might be a valid change, since these two activities are always executed at the start of their respective process version, and they are both exclusive to activity $c$.

Behavioral relations, as defined in \autoref{sec:behavioral_relations},
offer a formal framework for describing how activities relate to each other. In our technique, we use them to  algorithmically determine if two activities exhibit the same behavior, which would indicate that they likely implement the same process step. 
We first construct a relation matrix $R: A \times A \mapsto \{\rightarrow,\leftarrow,\#,\parallel\}$ for each event log. The behavioral relations that an activity has to all other activities in the same log (i.e., the rows of this matrix) are collectively referred to as its \emph{behavioral footprint}. In the next step, we then use these footprints as a basis to identify plausible replacements.

Deriving the behavioral relations between activities simply from the \emph{existence} of traces  (as done, e.g., in simple discovery algorithms \cite{alpha_miner}) is highly susceptible to noise. For example, when two activities only co-occur in a single trace out of thousands, they would be in a sequential relation according to the definition in \autoref{sec:behavioral_relations}, although that trace was likely incorrectly recorded and they are instead exclusive. Similar to, e.g., the inductive miner infrequent \cite{inductive_infrequent}, we therefore want to consider the frequency of traces when establishing behavioral relations. To this end, we compute scores for the exclusive and interleaving relation types based on their support in the log, i.e., the number of traces that exhibit the respective patterns.
We write $T_a$ for the set of traces in which $a$ occurs, $T_{a \land b}$ for the subset of $T_a$ in which both $a$ and $b$ occur, $T_{a \land \neg b}$ for the subset of $T_a$ in which $b$ does not occur, and $T_{a | a > b}$ for the subset of $T_a$ in which $a$ occurs before $b$.

\noindent The \textit{exclusiveness score} for $a$ and $b$ counts the number of traces that contain only one activity and normalizes it by the overall frequency of the activities: 
\vspace{-0.05em}
$$
s_{\#}(a, b) = min(\frac{|T_{a \land \neg b}|}{|T_a|}, \frac{|T_{b \land \neg a}|}{|T_b|})
$$

\noindent The \textit{interleaving order score} for $a$ and $b$ computes the difference between the number of traces in which $a$ is executed before $b$ and the number of traces in which $b$ is executed before $a$:

$$
s_{\parallel}(a, b) = 1 - \frac{abs(|T_{a | a > b}| - |T_{b | b > a}|)}{|T_{a \land b}|}
$$

\noindent After computing these scores, we first determine if $a$ and $b$ should be considered exclusive; this is the case if $s_{\#}(a, b) > \mathit{exc}$, where $\mathit{exc} \in [0,1]$ is a threshold parameter to be set by the user. If the activities are not exclusive, we determine if they should be considered interleaving in a similar manner: If $s_{\parallel}(a, b)$ is greater than the threshold parameter $\mathit{int} \in [0,1]$, $a$ and $b$ are interleaving. Both threshold parameters should be set close to 1, so that exclusive and interleaving relations are only inferred if the great majority of traces exhibit the respective patterns.
If there is not sufficient support for either an exclusivity or interleaving order relation, $a$ and $b$ are considered sequential in the direction that is more frequently observed, i.e., $a \rightarrow b$ if $|T_{a | a > b}| > |T_{a | b > a}|$ and $a \leftarrow b$ otherwise.

The output of this step is one matrix with pairwise activity relations for $L_1$ and $L_2$ each (see \autoref{fig:footprint_matching}). Each row vector in these matrices is one behavioral footprint of an activity. To differentiate between activity executions in the two logs, we use the notation $a_1$ and $a_2$ for the occurrence of activity $a$ in $L_1$ resp. $L_2$. Likewise, we denote the sets of activities in $L_1$ and $L_2$ as $A_1$ and $A_2$.

\subsection{Matching Activities between Logs}
\label{sec:matching}

Using the behavioral footprints from the previous step, we can identify activity replacements that are behaviorally plausible. Specifically, we consider an activity in $L_2$ to be a potential replacement for an activity in $L_1$ iff their behavioral footprints are identical. However, because we cannot assume $A_1 = A_2$---our assumption is merely that $A_1, A_2 \subseteq A$, given $A$ as our global set of activities---and the relation between two activities $a \in A_1$ and $x \notin A_1$ is undefined, it is not possible to establish equivalence based on the full footprint. Instead, we must disregard relations to activities that do not exist in both logs. Two footprints are then considered identical iff their relations to all activities shared between the two logs (i.e., to all activities in  the intersection $A_1 \cap A_2$) are identical. 

\begin{figure}[htb]
    \vspace{-2.0em}
    \centering
    \includegraphics[width=0.65\linewidth]{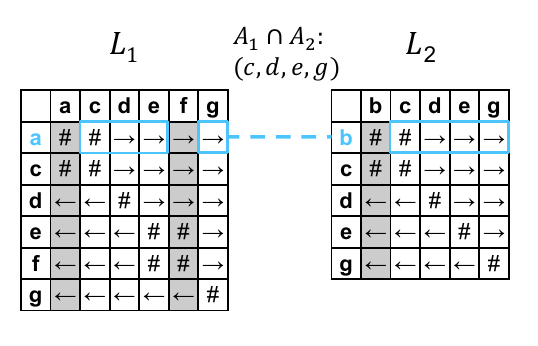}
    \caption{Matching activities between footprint matrices to find behaviorally plausible replacements in the running example.}
    \label{fig:footprint_matching}
    \vspace{-1.0em}
\end{figure}

\autoref{fig:footprint_matching} illustrates this for the running example: $a$, $b$, and $f$ are not in $A_1 \cap A_2$, so relations to them are not considered in the matching (indicated by greyed columns). $(a_1,b_2)$ is one potential match because their partial footprints (relations to $c$, $d$, $e$, and $g$, indicated in blue) are identical. In total, there are four behaviorally plausible replacements: $(a_1,b_2)$, $(a_1,c_2)$, $(c_1,b_2)$, and $(f_1,e_2)$.


The output of this step is a set of activity matches. A match $(a_1, b_2)$ implies that activity $a_1 \in A_1$ can be replaced with activity $b_2 \in A_2$. We remove any trivial matches (i.e., an activity matched with that same activity in the benchmark log, like $(a_1, a_2)$) from this set to get a set of potential replacement options.

\subsection{Combining Compatible Matches}
\label{sec:compatibility}

In the previous step, we have identified behaviorally feasible pairwise activity replacements, i.e., one activity being replaced by another one in isolation. 
In this step, we want to combine these isolated replacements into sets of process changes that can be implemented jointly. 
The rationale for this is that there may be dependencies between replacements, so that a certain replacement can only be implemented or will only lead to better performance, if it is combined with another one. 
In the running example, it might be the case that replacing $a$ with $b$ would improve performance in those traces that then execute $e$, but not in those that later continue with $f$. 
In this scenario, also implementing the replacement $(f_1, e_2)$ will result in an overall higher performance gain than only implementing $(a_1, b_2)$. We hence want to find sets of replacements that can later be jointly assessed for feasibility (Step 4) and performance impact (Step 5).

\begin{figure}[!ht]
        \vspace{-3.5em}
        \centering
        \begin{tikzpicture}[
            scale=0.85, every node/.style={scale=1},
            unanode/.style={circle, draw=black!75, minimum size=10mm, font=\bfseries}
            ]
            \node[unanode, draw=none]    (mab)    at(0,2)  {Matches:};
            \node[unanode, draw=none]    (mab)    at(0,1.5)  {$(a_1, b_2)$};
            \node[unanode, draw=none]    (mab)    at(0,1)  {$(a_1, c_2)$};
            \node[unanode, draw=none]    (mab)    at(0,0.5)  {$(c_1, b_2)$};
            \node[unanode, draw=none]    (mab)    at(0,0)  {$(f_1, e_2)$};
            \node[unanode]    (ab)    at(1,3)  {$\{a_1, b_2\}$};
            \node[unanode]    (ac)    at(5,3)  {$\{a_1, c_2\}$};
            \node[unanode]    (fe)    at(3,2)  {$\{f_1, e_2\}$};
            \node[unanode]    (cb)    at(3,0)  {$\{c_1, b_2\}$};
            \path [-, line width=0.5mm]  (fe) edge node[left] {} (ab);
            \path [-, line width=0.5mm]  (fe) edge node[left] {} (ac);
            \path [-, line width=0.5mm]  (fe) edge node[left] {} (cb);
            \path [-, line width=0.5mm]  (cb) edge node[left] {} (ab);
            \path [-, line width=0.5mm]  (cb) edge node[left] {} (ac);
            \node[unanode, draw=none]    (mab)    at(7.5,2)  {Max. Comp. Replacements:};
            \node[unanode, draw=none]    (mab)    at(7.5,1.5)  {$\{(a_1, b_2), (c_1, b_2), (f_1, e_2)\}$};
            \node[unanode, draw=none]    (mab)    at(7.5,1)  {$\{(a_1, c_2), (c_1, b_2), (f_1, e_2)\}$};
        \end{tikzpicture}
\caption{Matches, compatibility graph, and maximal compatible replacements from the running example. Fully connected subgraphs are also considered changes.}
\label{fig:compatibility_graph}
\vspace{-2.5em}
\end{figure}
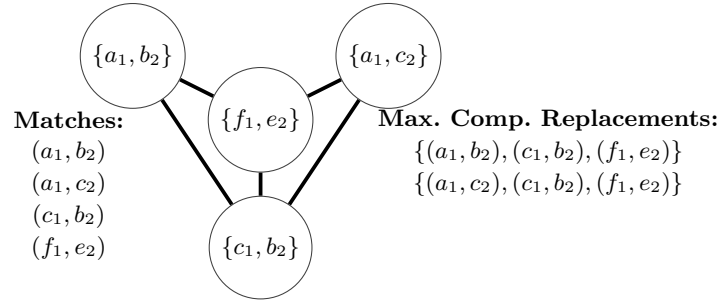

However, not all activity replacements can be freely combined with one another. 
Conflicts arise if the same activity in $L_1$ can be replaced by different activities from $L_2$. In the running example, both $(a_1, b_2)$ and $(a_1, c_2)$ are potential replacements, but only one of them can be implemented, so we should not show both to a process manager.
Therefore, we want to identify sets of changes that are not in conflict with each other and thus can be implemented together.
To do this, we construct a so-called compatibility graph (see \autoref{fig:compatibility_graph}), which contains one node for each possible replacement and an edge between two nodes iff they are not in conflict with one another, i.e., the two replacements do not refer to the same activity. 
The fully connected subgraphs of this compatibility graph (including single nodes) correspond to the sets of compatible replacements that are the output of this step. We call each set a potential \emph{process change}.

\subsection{Assessing the Feasibility of Process Changes}
\label{sec:feasibility_assessment}

At this stage, we have identified a set of behaviorally plausible process changes. Each process change $\delta$ consists of compatible replacement options between activities in $L_1$ and $L_2$. However, there is no guarantee that replacing activities with behaviorally similar ones will result in a feasible process. To ensure that the changes we propose can actually be implemented, we need to verify the feasibility of each process change. In the absence of process models, we base this verification on empirical evidence from the benchmark log $L_2$.

If we want to assess the feasibility of replacement option $(a_1, b_2)$, then for each variant $v_1 \in L_1$ in which $a$ is executed, we could check if $L_2$ contains a corresponding variant $v_2$ which is identical to $v_1$, except that $b$ is executed instead of $a$. If this is the case for all variants, we have strong evidence that replacing $a$ with $b$ is a valid change to make in the process. 
However, such exact matches are problematic because they do not account for differences in the benchmark process that are not part of the identified process changes, such as additional activities. They also do not consider the logs' potential incompleteness with regard to the permutations of concurrent or repeated activity executions. Therefore, instead of requiring exact matches of variants, we rely on trace alignments \cite{trace_alignments} to find those variants in the benchmark log that most closely match the ones from the own log and compute a \emph{feasibility score} for each process change.

For a single replacement $(a_1,b_2)$, let $V_{1,a} \subseteq L_1$ be the set of variants in $L_1$ where activity $a$ is executed and $V_{2,b}$ the set of variants in $L_2$ where activity $b$ is executed. We construct new variants in $L_1$, denoted $V'_{1,a}$, by implementing the replacement $(a_1, b_2)$ for each $v_1 \in V_{1,a}$. This yields a modified variant $v'_1$, where $a$ is replaced by $b$, but the other activities remain the same.
To assess the feasibility of this replacement, we examine each modified variant $v'_1$ and find its closest match among the variants in the benchmark log, i.e., the variant in $V_{2,b}$ that has the minimal edit distance (Levenshtein distance~\cite{editdistance}) to $v'_1$. We denote the closest match of $v'_1$ as $\mu(v'_1)$. We then iterate over all $v_1 \in V_{1,a}$ and compute the average frequency-weighted edit similarity between their modified versions $v'_1 \in V'_{1,a}$ and their closest matches $\mu(v'_1) \in V_{2,b}$:

$$
\text{Feasibility}(a_1,b_2) = \frac{\sum_{v_1 \in V_{1,a}} |T_1(v_1)| \cdot \text{EditSimilarity}(v'_1, \mu(v'_1))}{\sum_{v_1 \in V_{1,a}} |T_1(v_1)|}
$$

Using edit similarity instead of edit distance normalizes the feasibility score to $[0, 1]$. It is calculated by dividing the edit distance of two variants by the maximum length of these two variants, and then subtracting the result from $1$.

For a given process change $\delta$, which may consist of multiple replacement options, we assess the feasibility of a simultaneous implementation of all replacements in $\delta$ to the variants in $L_1$. For each variant $v_1 \in L_1$ in which any of the activities involved in $\delta$ are executed, we construct a modified variant $v'_1$ by applying all replacement options within $\delta$. For example, for the change $\delta_1 = \{ (a_1, b_2), (f_1, e_2) \}$ in our running example, there are three affected variants in $L_1$: $v_1 = \langle a, d, e, g \rangle$, $v_2 = \langle a, d, f, g \rangle$, and $v_3 = \langle c, d, f, g \rangle$. The modified variants after applying $\delta$ (replacing $a$ with $b$ and $f$ with $e$) would be $v'_1 = v'_2 = \langle b, d, e, g \rangle$ and $v'_3 = \langle c, d, e, g \rangle$. We then find the closest match for each variant and compute the feasibility score across all affected variants analogously to the single replacement case. In this example, each modified variant has an exact match, since $\langle b, d, e, g \rangle$ and $\langle c, d, e, g \rangle$ are recorded in $L_2$. The feasibility score for $\delta$ would therefore be $1$. For the process change $\delta_2 = \{ (a_1, b_2)$, meanwhile, the affected variants would be $v_1 = \langle a, d, e, g \rangle$ and $v_2 = \langle a, d, f, g \rangle$. In this case, the modified $v'_2 = \langle b, d, f, g \rangle$ does not have an exact match in $L_2$ and its closest alignment ($\langle b, d, e, g \rangle$) has an edit similarity of $0.75$. Assuming equal frequency of $v_1$ (which has an exact match) and $v_2$ (which has a closest match with similarity $0.75$), the overall feasibility score for this process change would be $0.875$.

\subsection{Assessing the Performance Impact of Process Changes}
\label{sec:performance_assessment}

The feasibility score from the previous section provides a means to assess whether it is \emph{possible} to implement a process change. To provide useful recommendations for process improvement, we also need to assess whether a process change will be \emph{beneficial}, i.e., have a (substantial) positive impact on process performance. We also base this assessment on empirical evidence from the benchmark event log: We approximate the expected performance impact of a process change as the difference in performance between the variants in $L_1$ that are affected by this change and their closest match variants in $L_2$.

For simplicity, we assume that both logs contain the same single performance measure $\pi(t)$ that assigns a value to each trace $t$, with higher values corresponding to better performance. The expected performance impact of a process change can then be approximated as the average difference in $\pi$ between the traces in $L_1$ where the to-be-replaced activities are executed and their corresponding (closest match) traces in $L_2$ where the replacing activities are executed. 

For a replacement $(a_1,b_2)$, we again construct modified variants from $V_{1,a}$ and find their closest match in $V_{2,b}$ in the benchmark log (see  \autoref{sec:feasibility_assessment}). For each $v_1$ in $V_{1,a}$, we calculate the average performance difference between all traces in $L_1$ of this variant ($T_1(v_1)$) and the traces in $L_2$ of its closest match $\mu(v'_1)$:

$$
\bar{\pi}(v_1) = \frac{\sum_{t \in T_1(v_1)} \pi(t)}{|T_1(v_1)|}.
$$

The expected performance impact across all variants affected by replacement $(a_1, b_2)$ is then calculated as the mean difference between the performance of traces that contain $a$ in $L_1$ and that of their closest match aligned traces in $L_2$:

$$
\Delta\pi(a_1, b_2) = \frac{\sum_{v_1 \in V_{1,a}} |T_1(v_1)| \cdot \left( \bar{\pi}(\mu(v'_1)) - \bar{\pi}(v_1) \right)}{\sum_{v_1 \in V_{1,a}} |T_1(v_1)|},
$$


\noindent For process changes that contain multiple replacements, we follow the same procedure as for the feasibility score: The expected performance impact of a process change $\delta$ is calculated with $V_{1, \delta}$ as the set of variants in $L_1$ in which at least one of the activities to be replaced in $\delta$ is executed and $V_{2,\delta}$ the set of variants in $L_2$ in which at least one of the replacement activities is executed. 

\subsection{End Result}
\label{sec:endresult}

Our technique outputs a list of process changes consisting of one or multiple activity replacement options, each associated with (1) a measure for feasibility, estimating the possibility to implement the change without further modifying the process and (2) a measure for the expected impact on process performance, indicating the potential benefits of the change. 
An exemplary output is shown in \autoref{tab:output}.
This list can be presented to a process manager, who can use the measures to filter, prioritize, and select the most promising changes for further investigation or implementation. Thereby, we offer evidence-based, prescriptive decision support in process improvement initiatives. 



\subsection{Discussion}

In designing our technique, we made certain decisions that resulted in limitations and could be addressed by potential extensions. 

\mypar{Focus on event logs}
We recommend process changes based exclusively on the information available in event logs, which capture only a fraction of the information about a process. For instance, a country branch may execute a certain activity due to regulatory requirements and cannot simply replace it, even if that would lead to a faster execution. Our technique might therefore yield false positive recommendations, which a manager must identify and discard.

\mypar{Concrete implementation}
We opted to prioritize simple and intuitive functions and measures for each step, which, of course, can be adapted. It would, e.g., be possible to extend the approach to consider multiple performance measures with individual weights or to use a utility function that considers feasibility and performance impact to derive a single quality score for each process change. It would also be possible to consider additional preprocessing steps, such as semantic matching in scenarios where activity names are not standardized.
    
\mypar{Similarity between logs}
If the process executions in the two logs are too dissimilar, our technique will not return any meaningful process changes. Consider two distinct processes that share only a start activity. In Step 2, all activities in $A_1$ would be matched with all activities in $A_2$, since they all share the same relation to the common start activity.
However, the only feasible process change in this scenario would be one that essentially replaces the entire own process with the benchmark one, which is not a useful recommendation.
The similarity between the logs can be established beforehand (e.g., \cite{earthmovers}), but there is no definite rule for how similar two logs must be to get good results.

\mypar{Focus on replacements}
Our technique is designed to identify activity replacements as practically relevant process modifications. Other possible modifications, such as adding an activity only found in the benchmark log or removing one that the benchmark does not execute, are not considered in this paper. However, the general ideas of our technique could be applied to these modifications as well.

\mypar{Behavioral matching}
In Step 1, we use behavioral relations to match activities. It would be possible to additionally consider semantics in this step, i.e., match activities based on the meaning of their labels.

\mypar{Single replacements}
In Step 2, we only consider 1:1 activity matches. Extending this to n:m matches could accommodate situations where replacements are only feasible when exchanging, e.g., entire choice constructs. This extension would significantly increase the complexity of our technique, as it necessitates identifying such constructs, even when they are nested. Process discovery algorithms address similar challenges and may be a useful starting point for this. 

    


\section{Evaluation}

Evaluating if our technique finds intended activity replacements requires pairs of event logs with performance measures and labeled differences. Since these are not available in real-life logs, we conduct an evaluation on synthetic data in \autoref{sec:experiment}.
To also show that our technique can be applied in real-life settings, we conduct a case study on a realistic event log in \autoref{sec:casestudy}.
All code, data, and parameters used in the experiment and the case study are available on GitLab.\footnote{\url{https://gitlab.uni-mannheim.de/jpmac/process-execution-benchmarking/}}

\subsection{Experiment}
\label{sec:experiment}

We used the Process and Log Generator (PLG) tool \cite{plg} to generate 1,000 randomly configured process models. We then created a corresponding benchmark process model for each of them by replacing 1-3 activities in the original model with new ones, keeping track of the replacements to establish a ground truth. In addition, we performed 0-2 activity insertions and deletions per model. These are not found by our technique, but increase the variance between the own and benchmark process and thus make the data more realistic.
We then simulated event logs from all models, with 1,000 traces each and control-flow noise, and applied our technique to the 1,000 own and benchmark event log pairs. 

We evaluated two aspects of our technique: (1) Whether the initial matches in Step 2 align with the ground truth (using precision and recall) and (2) whether the identified changes are more feasible than those returned by a random baseline. We constructed this baseline by randomly selecting the same number of matches from $A_1 \times A_2$ as the number of matches returned by the technique and then applying Steps 3 and 4 in the same way to both sets of matches. 
The results are shown in \autoref{tab:eval_results} (averaged over all log pairs). Precision and recall are relatively high, but do not approach $1$. The main reasons for this are incompleteness of either event log, particularly when it comes to larger blocks of concurrent or repeating activities, and noise, which can sometimes lead to incorrect activity relations being derived in Step 1. 
Overall, however, our technique identifies a great majority of intended matches and produces few false positives.

\begin{table}[bh]
    \vspace{-2em}
    \centering
    \caption{Mean precision, recall, and feasibility scores achieved in the evaluation.}
    \label{tab:eval_results}
    \vspace{0.5em}
    \begin{tabular}{llcc}
    \toprule
        Aim & Metric & Technique & Baseline \\
        \midrule
       (1) & Precision  &  0.831 & - \\
        & Recall  & 0.901 & - \\
       \midrule
      (2) & Feasibility Score & 0.802 & 0.580 \\
       \bottomrule
    \end{tabular}
    \vspace{-1em}
\end{table}

The average feasibility score is also considerably higher than for the random baseline. This indicates that when the process changes our technique identifies are implemented, the resulting traces are likely to have equivalents with low edit distance in the benchmark logs and therefore be valid process executions. Note that because of the way we have set up this evaluation, feasibility scores close to $1$ are unlikely to occur, because the activity insertions and deletions inherently introduce disparities between otherwise corresponding variants. 

Evaluating the performance assessment would require simulating the logs such that the traces resulting from specific replacements in the benchmark log demonstrate better performance than their matches in the original log and then showing that our technique assigns good performance measures to them. By design, our approach involves identifying these exact traces in the benchmark log through alignments and calculating a performance difference to the original ones (see \autoref{sec:performance_assessment}). This evaluation would therefore not be informative w.r.t. its efficacy. Consequently, we omit an explicit evaluation in this experiment.

\subsection{Case study}
\label{sec:casestudy}

To showcase how our technique can be applied in real-life settings and what the recommended changes could look like, we also conduct a small case study on a realistic event log. This event log consists of demo data that was released as part of the SAP Signavio Plug and Gain initiative; it simulates typical executions of a purchasing process in the SAP ERP system. 
We split this event log into two parts that each correspond to a country-level suborganization. The log of the India organization has a median case duration of about 11 days and acts as the ``own'' event log in this scenario. The log of the Germany organization, which acts as our benchmark, has a median case duration of about 9 days.

We filter out the 5\% least common variants from this event log and then apply our technique. In doing so, we identify five potential process changes that all pertain to the creation of the initial purchase order (\autoref{tab:output}). We can see that, in this dataset, all process changes have a high feasibility, i.e., when the changes would be implemented in the own process, all frequently observed variants have exact matches in the benchmark dataset. However, there are stark differences in their expected performance impact (measured in average hours lost or gained per case): Only the first two changes would lead to a reduction of about 1 and 4 hours in average throughput time across all cases, whereas the remaining three are associated with an increase of over 8 hours.
A straightforward conclusion from these results would be that creating a PO in the SCM leads to considerably slower process executions, and that creating it directly using the "Create PO" dialog window is a faster option.

\begin{table}[htb]
    \vspace{-2em}
    \caption{Exemplary list of process changes for the running example}
    \label{tab:output}
    \centering
    \begin{tabularx}{\linewidth}{llcc}
    \toprule
       Activities & Replacements & Feasibility & Performance \\
        \midrule
        Create PO item from SCM  & Convert PR item into PO item  & $0.95$ & $-0.97$ h/case\\
        \midrule
        Create PO item from SCM & Create PO item in dialog & $0.95$ & $-4.42$ h/case \\
        \midrule
        Create PO item in dialog  & Convert PR item into PO item  & $0.96$ & $+11.61$ h/case\\
        \midrule
        Create PO item from SCM & Convert PR item into PO item & $0.96$ & $+8.96$ h/case\\
        Create PO item in dialog & Convert PR item into PO item & &  \\
        \midrule
        Create PO item from SCM & Create PO item in dialog & $0.96$ & $+8.24$ h/case\\
        Create PO item in dialog & Convert PR item into PO item & &  \\
        \bottomrule
    \end{tabularx}
\end{table}

\section{Related Work}


\mypar{Process Performance Improvement}
Improving process performance is the end goal of most process mining initiatives \cite{Stein2024_improvement}, and several studies focus on the performance aspects of event data. Some explore the integration of performance measures with standard process mining analyses, such as process discovery \cite{performance_discovery1}, conformance checking \cite{performance_conformance1}, and process simulation \cite{simulation_performance}. Others aim to analyze the performance of running process instances, e.g., by predicting the remaining time of a case \cite{timeprediction} or prescribing interventions to avoid negative process outcomes \cite{prescriptive}.

Another related literature stream investigates how to make process mining results actionable, i.e., move from understanding a process to being able to improve it \cite{Stein2024_improvement}. This includes technical papers introducing new process mining approaches \cite{Park2022_actionable,action_oriented_2,action_oriented3} as well as case studies and qualitative research about how practitioners utilize process mining results \cite{performance_industry1,performance_industry2,Stein2024_improvement}.

\mypar{Process Variant Analysis}
There is substantial process mining literature on analyzing and comparing different versions or variants of the same process (e.g., \cite{pva1,pva2,pva7,pva8}; see \cite{process_variants_review} for a comprehensive review). The focus in these publications is on (1) identifying and (2) visualizing differences between two or more process variants, but not on associating these differences with performance implications or making recommendations for process improvement \cite{process_variants_review}. Most similar to our work is \cite{process_benchmarking}, in which event logs from different organizations are initially grouped based on their overall performance, as measured by key performance indicators, and then compared. However, this approach remains descriptive and does not suggest concrete process modifications.

\mypar{Process Model Matching}
While identifying activity correspondences among separate event logs for process execution benchmarking is a novel topic, there has been substantial research on finding these correspondences in process models. This task, called process model matching \cite{pm_matching,pm_matching_semantic1}, generally has a different objective than process execution benchmarking: Its main purpose is to align or integrate handcrafted process models, whereas execution benchmarking aims to derive process improvement opportunities. Notably, process model matching can also be used for the purpose of process improvement when employed with a reference model \cite{pm_matching_benchmarking}, but this is not its typical application.


\section{Conclusion}

In this paper, we have introduced a novel technique for automated process improvement through process execution benchmarking. It relies on behavioral relations to identify potential activity replacements using a benchmark event log and associates them with measures for feasibility and expected performance impact. Our technique can supplement high-level indicator-based process benchmarking by leveraging event logs to recommend to a process manager concrete, actionable process modifications that are expected to improve performance. 

In future work, one could explore technical extensions to the approach presented. It would also be worthwhile to conduct a user study with process managers to verify if the outputs are useful for process improvement in practice.

\bibliographystyle{splncs04}
\bibliography{bibliography}

\end{document}